\documentclass[11pt,a4paper]{article}
\pdfoutput=1

\usepackage[utf8]{inputenc}
\usepackage[T1]{fontenc}        
\usepackage{amsmath,amssymb,amsfonts}
\usepackage{graphicx}
\usepackage{booktabs}
\usepackage{multirow}

\usepackage[numbers,sort&compress]{natbib}

\usepackage{hyperref}
\usepackage{cleveref}
\usepackage{xcolor}
\usepackage{algorithm}
\usepackage{algpseudocode}
\usepackage{subcaption}
\usepackage[margin=2.5cm]{geometry}
\usepackage{bm}
\usepackage{tikz}
\usetikzlibrary{positioning,arrows.meta,calc}

\hypersetup{
    colorlinks=true,
    linkcolor=blue,        
    citecolor=blue,        
    urlcolor=blue,         
    pdftitle={Sparse Mamba Decoder for Quantum Error Correction},
    pdfauthor={Sayedsalehi, Bagherzadeh, Shcherbakov, Gaudiot}
}

\newcommand{\dmodel}{d_{\mathrm{model}}}
\newcommand{\dstate}{d_{\mathrm{state}}}
\newcommand{\dconv}{d_{\mathrm{conv}}}
\newcommand{\dread}{d_{\mathrm{read}}}
\newcommand{\lres}{L_{\mathrm{res}}}
\newcommand{\wgate}{w_{\mathrm{gate}}}
\newcommand{\bigO}{\mathcal{O}}

\definecolor{mwpmblue}{HTML}{377EB8}
\definecolor{tessred}{HTML}{E41A1C}
\definecolor{mambagreen}{HTML}{4DAF4A}

\title{Sparse Mamba Decoder for Quantum Error Correction:\\
Efficient Defect-Centric Processing of Surface Code Syndromes}

\author{
  Samira Sayedsalehi\thanks{Corresponding author. Email: ssayedsa@uci.edu},
  Nader Bagherzadeh,
  Maxim Shcherbakov, 
  and Jean-Luc Gaudiot \\[6pt]
  \textit{Department of Electrical Engineering and Computer Science} \\
  \textit{University of California, Irvine, CA 92697, USA}
}
\date{}

\begin{document}
\maketitle


\begin{abstract}
Quantum error correction (QEC) is essential for building fault-tolerant 
quantum computers, requiring decoders that are simultaneously accurate, 
fast, and scalable. Most state-of-the-art neural decoders achieve high 
accuracy but process the full dense syndrome array of size 
$\bigO(d^2 R)$ regardless of the actual error rate, where $d$ is the 
code distance and $R$ is the number of measurement rounds. At physically 
relevant error rates ($p \approx 0.1\%$), fewer than 5\% of syndrome 
entries contain active detection events---yet existing decoders process 
the entire syndrome volume. We introduce the Sparse Mamba Decoder (SMD), 
a defect-centric neural decoder that processes only the $k$ active 
detection events using a 13-dimensional feature representation per 
defect and a Mamba state-space backbone, achieving $\bigO(k)$ complexity. 
Across depolarizing, uniform circuit-level, SI1000, and Google Sycamore 
experimental benchmarks, SMD reduces the MWPM logical error rate by up 
to 49\% at $d \leq 5$ under SI1000 noise, runs 95--467$\times$ faster 
than the Tesseract near-MLD decoder and 232--463$\times$ faster than 
Belief Matching, and maintains nearly constant latency 
(24--57\,$\mu$s) across $d = 3$--$9$ under uniform circuit-level noise. 
On the Sycamore experimental dataset, the SMD ensemble matches or 
slightly surpasses the dense Mamba decoder of Lee et al. All 
results are obtained on commodity NVIDIA GPUs with $7.5$--$16$M 
parameters, without specialized accelerators.
\end{abstract}

\section{Introduction}
\label{sec:introduction}

Fault-tolerant quantum computing requires logical error rates far below 
those achievable with physical qubits, necessitating quantum error 
correction (QEC) codes that encode logical information redundantly across 
many physical qubits~\cite{shor1995scheme,kitaev2003fault}. The surface 
code~\cite{dennis2002topological,fowler2012surface} is the leading 
candidate for near-term implementation due to its high threshold error rate 
and local stabilizer structure, and continues to see active research 
in code design~\cite{sayedsalehi2025defect} and decoder development. However, the practical utility of any QEC 
code depends critically on the \emph{decoder}---the classical algorithm 
that infers logical errors from noisy stabilizer measurements.

An ideal decoder must satisfy three competing requirements simultaneously: 
\emph{accuracy} (approaching the maximum-likelihood estimate), \emph{speed} 
(keeping pace with the quantum hardware cycle time, typically 
${\sim}1\,\mu$s for superconducting qubits~\cite{google2023suppressing}), 
and \emph{scalability} (maintaining performance as code distance increases). 
In practice, existing decoders make tradeoffs among these three axes, and 
achieving all three simultaneously---particularly at the code distances 
required for fault tolerance---remains an open challenge.

Minimum-Weight Perfect Matching (MWPM)~\cite{higgott2023sparse, higgott2022pymatching} 
is fast and scalable but suboptimal, as it does not account for correlated 
errors. The Tesseract decoder~\cite{beni2025tesseract} achieves 
near-optimal accuracy through beam search over the detector error model 
but requires millisecond-scale computation. 
AlphaQubit~\cite{bausch2024learning} introduced a recurrent-transformer 
architecture that outperforms all matching-based decoders on experimental 
data, but its $\bigO(d^4)$ per-round attention complexity  makes real-time decoding 
impractical at large distances. AlphaQubit~2~\cite{alphaqubit2} addresses 
this with a redesigned architecture achieving $\bigO(d^2)$ per-round scaling and 
real-time decoding up to distance 11, but requires specialized 
Trillium TPU hardware~\cite{vahdat2024trillium}. Lee et 
al.~\cite{lee2025scalable} proposed replacing the transformer with a 
Mamba state-space model~\cite{gu2024mamba}, reducing the complexity to 
$\bigO(d^2)$ while maintaining accuracy on consumer GPUs.

Despite these advances, most existing neural decoders share a 
fundamental inefficiency: they process the \emph{full} syndrome array 
of size $d^2 \times R$ at every decoding step, regardless of how many 
errors actually occurred. The graph-neural-network decoder of Lange et al.~\cite{lange2023gnn} 
operates on detection events but requires explicit graph construction 
and message-passing operations that are less GPU-friendly than 
pure sequence processing. At a physical error rate of $p = 10^{-3}$ under SI1000 noise, only a few percent of stabilizer measurements fire---we measure detection-event 
densities of 2.8--3.2\% across $d = 3$--$7$ (Table~\ref{tab:sparsity}). 
Yet existing decoders process every entry of the syndrome volume, 
analogous to running a dense matrix multiplication on a 97\%-sparse matrix.

In this work, we introduce the Sparse Mamba Decoder (SMD), which 
fundamentally changes the input representation from dense syndrome arrays 
to sparse defect sequences. Each active detection event is represented by 
a 13-dimensional feature vector encoding spatial coordinates on the 
rotated lattice, stabilizer type ($X$ or $Z$), spatial and temporal 
neighborhood connectivity flags, normalized distances to the logical 
boundaries, and a reconstructed stabilizer measurement computed via 
cumulative XOR of the per-stabilizer detection event history. The 
resulting variable-length sequences---typically $k \ll d^2 R$ at 
physically relevant error rates---are processed by a Mamba state-space 
backbone~\cite{gu2024mamba} in linear time without attention. This yields 
$\bigO(k)$ complexity, compared to $\bigO(d^2 R)$ for dense neural 
decoders and $\bigO(d^4 R)$ for attention-based methods. Throughout 
this paper, complexity is reported per decoding task (one logical qubit 
memory experiment over $R$ rounds); equivalent per-round costs for 
recurrent decoders are obtained by dividing by $R$, e.g., $\bigO(d^2)$ 
per round for dense Mamba~\cite{lee2025scalable} and $\bigO(d^4)$ 
per round for AlphaQubit~\cite{bausch2024learning}.

Our main contributions are:
\begin{enumerate}
    \item \underline{A sparse defect-centric architecture for QEC decoding.} 
    SMD processes only active detection events through a Mamba 
    state-space backbone, achieving $\bigO(k)$ complexity that scales 
    with the number of errors rather than the code size. The 
    13-dimensional defect features encode the spatial structure and 
    measurement history needed for accurate decoding without explicit 
    2D convolutions.
    \item \underline{Strong accuracy across synthetic and experimental 
    benchmarks.} Under SI1000 noise at $p = 1.5 \times 10^{-3}$, SMD 
    reduces the MWPM logical error rate by up to 49\% at $d \leq 5$ 
    and by 16\% at $d = 7$ via a three-checkpoint ensemble, where a 
    Libra-style ensemble of seven perturbed MWPM decoders gives no 
    improvement, isolating the gain to learned correlations. On the 
    Sycamore dataset, the SMD ensemble matches or slightly surpasses 
the dense Mamba decoder of Lee et al.~\cite{lee2025scalable} 
at both $d = 3$ and $d = 5$, while the single-model SMD remains 
competitive at $d = 3$ and approaches dense-Mamba performance 
at $d = 5$.
    \item \underline{Order-of-magnitude speedups over near-MLD decoders.} 
    Under SI1000 noise, SMD runs 95$\times$--467$\times$ faster than 
    Tesseract and 232$\times$--463$\times$ faster than Belief Matching, 
    with the speedup growing with code distance. Under uniform 
    circuit-level noise, decoding latency is nearly constant 
    (24--57\,$\mu$s) across $d = 3$--$9$, compared to a 773$\times$ 
    increase for Belief Matching over the same range.
    \item \underline{Commodity-hardware feasibility.} All results are 
    obtained on commodity NVIDIA GPUs---an RTX 4090 (24\,GB) for the 
    smaller depolarizing and uniform circuit-level models, and a single 
    H200 NVL (141\,GB) for the larger SI1000 model---without the 
    specialized Trillium TPUs used by 
    AlphaQubit~2~\cite{alphaqubit2,vahdat2024trillium}.
\end{enumerate}

\section{Background}
\label{sec:background}

\subsection{Surface Codes and Stabilizer Measurements}
\label{sec:surface_codes}

\begin{figure}[ht!]
    \centering
    \includegraphics[width=0.95\linewidth]{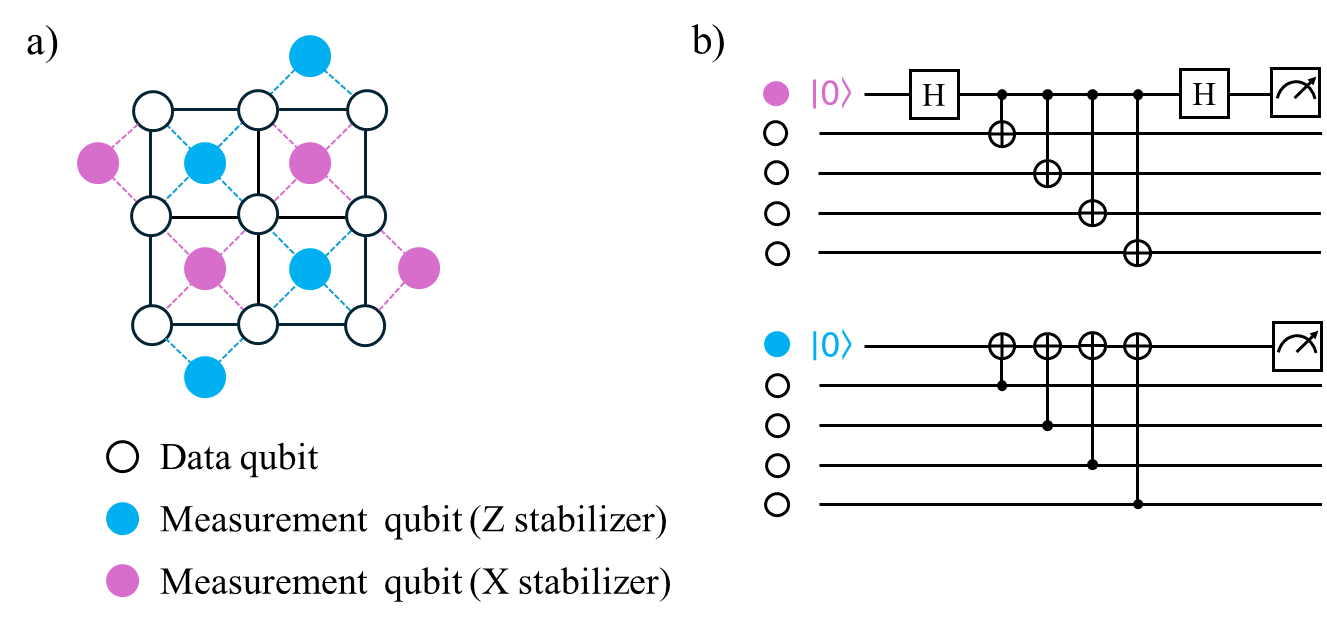}
    \caption{a) Planar layout of a rotated surface code with code distance
    $d = 3$. White circles represent data qubits, while blue circles
    \textcolor{blue}{$\bullet$} and pink circles \textcolor{magenta}{$\bullet$}
    represent $Z$-type and $X$-type stabilizer measurement qubits,
    respectively. b) Circuit representation of the $X$-type (top) and
    $Z$-type (bottom) stabilizer measurements. Open circles $(\circ)$ denote
    the four data qubits coupled to each measurement qubit.}
    \label{fig:surface_code}
\end{figure}

The rotated surface code of distance $d$ encodes one logical qubit into
$d^2$ data qubits using a CSS (Calderbank-Shor-Steane) structure of
$(d^2 - 1)$ stabilizer measurements arranged in a checkerboard pattern on
a two-dimensional lattice~\cite{fowler2012surface}
(Figure~\ref{fig:surface_code}). These stabilizers form two independent
families: $Z$-type checks detect $X$-errors (bit flips) by measuring the
$Z$-parity of adjacent data qubits, while $X$-type checks detect $Z$-errors
(phase flips) by measuring the $X$-parity of their neighbors; a $Y$-error
triggers both check types simultaneously, enabling its identification. In a
memory experiment, stabilizer measurements are repeated for $R$ rounds,
producing a syndrome array of shape $(d^2 - 1) \times R$, with $X$ and $Z$
corrections handled independently.

A \emph{detection event} at stabilizer $i$ and round $t$ is defined as the
temporal difference of consecutive measurements:
\begin{equation}
    d_{i,t} = s_{i,t} \oplus s_{i,t-1},
    \label{eq:detection_event}
\end{equation}
where $s_{i,t} \in \{0, 1\}$ is the raw measurement outcome and $\oplus$
denotes XOR\@. In the absence of errors, all detection events are zero;
physical errors on data qubits trigger clusters of nearby detection events
in space-time.

The decoder's task is to predict the logical observable
$\lambda \in \{0, 1\}$ from the set of detection events
$\{d_{i,t}\}_{i,t}$, determining whether a logical error has occurred.

\subsection{The Decoding Problem}
\label{sec:decoding_problem}

Given a syndrome $\sigma = \{d_{i,t}\}$, the optimal decoder predicts 
the most likely logical observable:
\begin{equation}
\hat{\lambda} = \arg\max_{\lambda \in \{0,1\}} P(\lambda \mid \sigma),
\end{equation}
which requires marginalizing over all physical error configurations 
consistent with the observed syndrome. This decoding problem is 
NP-hard in general~\cite{iyer2015hardness}. Practical decoders 
approximate this inference using matching-based methods 
(MWPM~\cite{higgott2023sparse}), tensor-network 
contraction~\cite{bravyi2014efficient}, or learned neural 
networks~\cite{bausch2024learning,lee2025scalable}.

\subsection{Related Work}
\label{sec:related_work}
 
Matching-based decoders reduce the decoding problem to a 
minimum-weight perfect matching on a graph derived from the detector 
error model. MWPM~\cite{higgott2022pymatching} and its correlated 
variants achieve a circuit-level threshold 
of approximately 0.5--1\%, with Sparse Blossom~\cite{higgott2023sparse} 
achieving $\bigO(n^{1.32})$ observed complexity. 
Union-Find~\cite{delfosse2021almost} offers near-linear 
$\bigO(n\alpha(n))$ complexity at slightly reduced accuracy. Belief 
Matching~\cite{higgott2022beliefmatching} combines belief propagation 
with matching-based decoding and improves accuracy under circuit-level 
noise, although belief propagation alone can perform poorly on surface 
codes due to short cycles in the Tanner graph. The Tesseract 
decoder~\cite{beni2025tesseract} reaches near-optimal accuracy 
through beam search over the detector error model, but at 
millisecond-scale per-shot cost. Recently, ensemble approaches such as 
Libra~\cite{jones2024libra} have explored combining multiple perturbed 
matching decoders to close the residual gap to maximum-likelihood 
decoding. For small codes, contracting the tensor network defined by 
the syndrome distribution provides a near-optimal but exponentially 
expensive baseline~\cite{bravyi2014efficient}, which we use as a 
reference on the Sycamore experimental dataset.
 
Neural decoders learn representations of the syndrome history from data, reducing reliance on explicit error-model priors. AlphaQubit~\cite{bausch2024learning} 
introduced a recurrent-transformer architecture trained on both 
simulated and experimental data, outperforming matching decoders on 
the Sycamore quantum processor. Its successor, 
AlphaQubit~2~\cite{alphaqubit2}, includes a compact real-time variant 
(AQ2-RT) that achieves $<1\,\mu$s/cycle decoding up to distance 11 
on Trillium TPUs~\cite{vahdat2024trillium}. Lee et 
al.\cite{lee2025scalable} replaced the transformer core with a 
Mamba state-space model~\cite{gu2024mamba}, reducing complexity from 
$\bigO(d^4)$ to $\bigO(d^2)$ per round while matching transformer 
accuracy on the Sycamore benchmark with logical error rates of 
$2.98\%$ at $d{=}3$ and $3.03\%$ at $d{=}5$. Lange et 
al.~\cite{lange2023gnn} proposed a graph-neural-network decoder for 
circuit-level noise, with performance approximately matching MWPM. 
These dense neural decoders process information proportional to the 
full syndrome volume $d^2 \times R$, regardless of the actual error 
rate. In contrast, our Sparse Mamba Decoder processes only the $k$ 
active detection events, yielding neural-processing cost $\bigO(k)$ 
with $k \ll d^2 R$ at physically relevant error rates. 
Table~\ref{tab:decoder_comparison} summarizes the comparison.


\section{Method: Sparse Mamba Decoder}
\label{sec:method}

\begin{figure}[ht!]
    \centering
    \includegraphics[width=\linewidth]{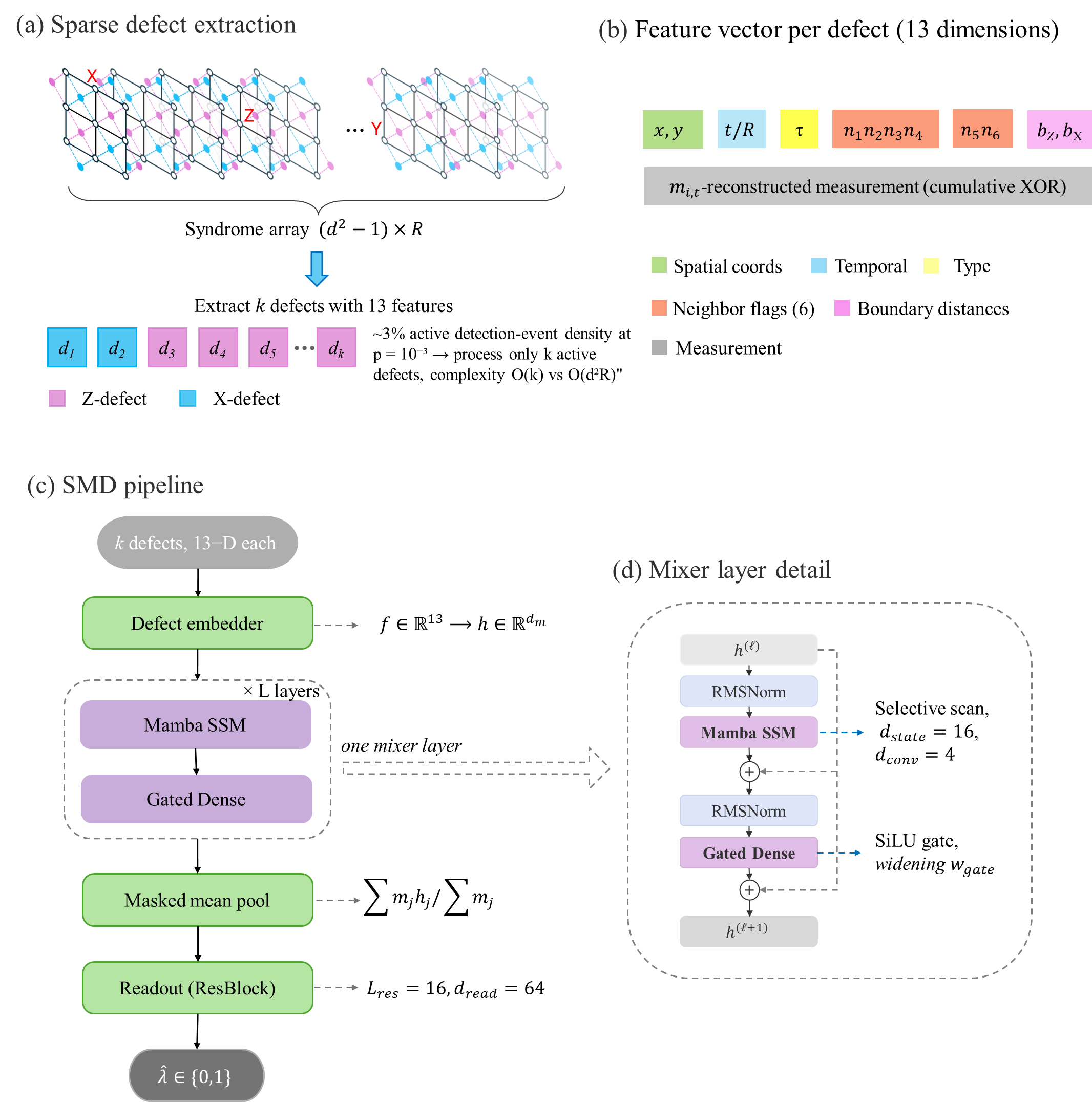}
    \caption{Sparse Mamba Decoder architecture.
    \textbf{(a)} Sparse defect extraction from a $(d^2{-}1) \times R$
    syndrome volume to $k$ defect tokens $d_1, \ldots, d_k$
    ($k \ll d^2 R$ at physically relevant error rates).
    \textbf{(b)} 13-dimensional feature vector per defect: spatial
    coordinates $(x, y)$, normalized time $t/R$, stabilizer type $\tau$,
    spatial and temporal neighbor flags, boundary distances
    $b_Z, b_X$, and the reconstructed measurement $m_{i,t}$ from
    cumulative XOR.
    \textbf{(c)} Pipeline: per-token embedder
    $\mathbb{R}^{13} \to \mathbb{R}^{\dmodel}$, $L$ mixer layers,
    masked mean pool, ResBlock readout to logical observable
    $\hat{\lambda}$.
    \textbf{(d)} One mixer layer: RMSNorm $+$ Mamba selective scan
    ($\dstate{=}16$, $\dconv{=}4$), then RMSNorm $+$ Gated Dense
    (SiLU gating, widening $\wgate$), both with residual connections.}
    \label{fig:architecture}
\end{figure}

A physical Pauli error on a data qubit persists until corrected; left in
place, it would cause its adjacent stabilizers to fire every subsequent
round, producing a dense ``always-on'' syndrome. The detection event
$d_{i,t} = s_{i,t} \oplus s_{i,t-1}$ converts this persistent state into
a sparse spacetime signal: each error contributes detection events only
at the \emph{endpoints} of its trajectory in spacetime, where it appears
and where it is cancelled. This is the structural reason the syndrome
volume is overwhelmingly zero at physical error rates of interest.
 
For a distance-$d$ surface code with $R$ rounds of stabilizer
measurements, the syndrome volume contains $(d^2 - 1) \times R$ entries,
but the expected number of active detection events is
\begin{equation}
    \mathbb{E}[k] \approx (d^2 - 1) \times R \times \rho(p),
    \label{eq:expected_defects}
\end{equation}
where $\rho(p)$ is the detection event density at physical error rate
$p$. At $p = 10^{-3}$ under SI1000 noise, we measure $\rho \approx 0.03$
(Table~\ref{tab:sparsity}), yielding $k \ll d^2 R$
(${\sim}30\times$ reduction at $d = 7$, $R = 120$). SMD operates on this
$k$-token sparse sequence, not on the full $d^2 R$ volume
(Fig.~\ref{fig:architecture}a).
 
We extract the active detection events from the entire syndrome volume
and represent each as a 13-dimensional feature vector
(Fig.~\ref{fig:architecture}b). For a defect at
stabilizer $i$ in round $t$, this vector is
\begin{equation}
    \mathbf{f}_{i,t} = \big[\underbrace{x_i,\, y_i}_{\text{spatial}},\;
    \underbrace{t / R}_{\text{temporal}},\;
    \underbrace{\tau_i}_{\text{type}},\;
    \underbrace{n_1,\ldots,n_4}_{\text{spatial nb.}},\;
    \underbrace{n_5,\, n_6}_{\text{temporal nb.}},\;
    \underbrace{b_Z,\, b_X}_{\text{boundary}},\;
    \underbrace{m_{i,t}}_{\text{measurement}}\big],
    \label{eq:features}
\end{equation}
where $(x_i, y_i)$ are the normalized stabilizer coordinates on the
rotated lattice; $t/R$ is the normalized round index;
$\tau_i \in \{0, 1\}$ indicates the stabilizer type (X- or Z-type);
$n_1, \ldots, n_4$ are binary flags indicating whether each of the four
spatial neighbors also fired in round $t$; $n_5, n_6$ are binary flags
for the temporal neighbors at rounds $t-1$ and $t+1$; $b_Z, b_X$ are
normalized distances to the Z and X logical boundaries; and $m_{i,t}$ is
the reconstructed measurement, computed as the cumulative XOR of
detection events:
\begin{equation}
    m_{i,t} = d_{i,1} \oplus d_{i,2} \oplus \cdots \oplus d_{i,t}
    = \bigoplus_{s=1}^{t} d_{i,s}.
    \label{eq:cumxor}
\end{equation}
The reconstructed measurement $m_{i,t}$ recovers the raw stabilizer
measurement outcome (up to the unknown initial state), providing the
model with a dual view: the detection event indicates \emph{where errors
changed}, while the reconstructed measurement indicates the \emph{current
state} of each stabilizer. This dual-input approach was shown to be
beneficial by Bausch et al.~\cite{bausch2024learning} for dense
representations; we adapt it here for sparse defect sequences.
 
After all $R$ rounds of stabilizer measurement complete, we extract the
active defects from the entire syndrome volume in one pass and sort them
by time (primary) and stabilizer index (secondary). This produces a
single ordered defect sequence of length $k$. Sequences are padded to a
fixed maximum length $k_{\max}$ with a binary mask, yielding a batch
representation of shape $(B, k_{\max}, 13)$ with mask $(B, k_{\max})$.
Critically, defects are processed as a unified sequence: SMD does not
advance a hidden state (the internal vector representation
$\mathbf{h}^{(\ell)} \in \mathbb{R}^{\dmodel}$ carrying information
through the network) per measurement round. While the one-time
dense-to-sparse extraction requires a pass over the syndrome volume
(cost $\bigO(d^2 R)$), all subsequent neural processing scales as
$\bigO(k)$ in the active defect count. Future hardware-integrated
implementations could emit detection events directly as a sparse
stream, eliminating the dense scan and making the entire pipeline
sparse.

\subsection{Decoder Architecture}
\label{sec:architecture}
 
SMD processes the $k$-token defect sequence in a single forward pass
(Fig.~\ref{fig:architecture}c):
\begin{equation}
    \text{DefectEmbed} \;\to\;
    L \times \text{MixerLayer} \;\to\;
    \text{MaskedPool} \;\to\; \text{Readout} \;\to\; \hat{\lambda}.
\end{equation}
There is no per-round recurrence: a single Mamba stack consumes the
entire defect sequence at once.
 
The defect embedder projects the 13-dimensional feature vector to the
model dimension $\dmodel$ through a two-layer MLP with LayerNorm and
GELU activation, applied independently and in parallel to each token:
\begin{equation}
    \mathbf{h}^{(0)} = \text{LayerNorm}\big(W_2 \,\text{GELU}(\text{LayerNorm}(W_1 \mathbf{f} + b_1)) + b_2\big).
\end{equation}
 
Each of the $L$ subsequent mixer layers applies a Mamba selective
state-space block~\cite{gu2024mamba} followed by a gated dense block,
with RMSNorm and residual connections (Fig.~\ref{fig:architecture}d):
\begin{align}
    \mathbf{z} &= \mathbf{h}^{(\ell)} + \text{Mamba}\big(\text{RMSNorm}(\mathbf{h}^{(\ell)})\big), \\
    \mathbf{h}^{(\ell+1)} &= \mathbf{z} + \text{GatedDense}\big(\text{RMSNorm}(\mathbf{z})\big),
\end{align}
where the GatedDense block consists of two parallel linear projections
with SiLU gating:
\begin{equation}
    \text{GatedDense}(\mathbf{x}) = W_c \big(\text{SiLU}(W_a \mathbf{x}) \odot W_b \mathbf{x}\big).
\end{equation}
The widening factor $\wgate$ controls the intermediate dimension
$\wgate \cdot \dmodel$. The Mamba block uses a selective scan
mechanism~\cite{gu2024mamba} with state dimension $\dstate$ and
convolution width $\dconv$. The Mamba block mixes information
\emph{across} the $k$ tokens of the defect sequence, while the
GatedDense block mixes information \emph{within} each token across the
$\dmodel$ channels.
 
In contrast to the dense recurrent decoders of
AlphaQubit~\cite{bausch2024learning,alphaqubit2} and Lee et
al.~\cite{lee2025scalable}, our mixer layer contains no explicit
spatial-reshape operations: the Syndrome Mixer of lee et al.\
additionally requires Scatter-to-2D, dilated 2D convolutions, and
Gather-from-2D operations to capture geometric correlations across
stabilizers, whereas SMD encodes spatial structure directly in the 13-D
feature vector and processes the defect sequence with Mamba alone.
 
After $L$ mixer layers, we aggregate the variable-length defect
representations using masked mean pooling:
\begin{equation}
    \bar{\mathbf{h}} = \frac{\sum_{j} m_j \cdot \mathbf{h}^{(L)}_j}
    {\sum_{j} m_j + \epsilon},
\end{equation}
where $m_j \in \{0, 1\}$ is the binary mask for position $j$ (1 for real
defects, 0 for padding). The pooled representation is then passed
through a readout network that produces the logit
$z \in \mathbb{R}$, which maps to the predicted probability
$P(\lambda = 1 \mid \sigma) = \sigma(z) = 1/(1 + e^{-z})$. The 
predicted observable is $\hat{\lambda} = \mathbb{I}[\sigma(z) > 0.5]$. 
We employ two readout variants: a ResBlock readout consisting of 
$\lres$ residual blocks with hidden dimension $\dread$, and a simpler 
MLP readout with a single hidden layer.
 
For depolarizing noise with perfect stabilizer measurements
(Section~\ref{sec:depolarizing}), we use a dual-head decoder that
predicts both the Z-type logical observable $\lambda_Z$ and
the X-type logical observable $\lambda_X$ from a shared backbone
representation, following the setup of Lange et
al.~\cite{lange2023gnn}.

\subsection{Training Protocol}
\label{sec:training_protocol}
 
We employ three training protocols depending on the noise model and
data source.
 
For uniform circuit-level and depolarizing noise benchmarks, we train
from scratch with fresh synthetic data generated each epoch. The
physical error rate is sampled uniformly from a training set (e.g.,
$p \in \{0.001, 0.002, \ldots, 0.005\}$) following Lange et
al.~\cite{lange2023gnn}. We use AdamW with cosine learning rate
annealing over 500--1000 epochs, with $4$--$6 \times 10^6$ freshly
generated samples per epoch.
 
For the SI1000 benchmark, we train with a three-stage curriculum
over measurement rounds: the maximum round count increases from
$r \leq 9$ (Stage~1) to $r \leq 25$ (Stage~2) to $r \leq 120$
(Stage~3) at transitions of 500K and 1M iterations. The physical error rate is sampled uniformly from
$p \in \{0.001, 0.002, 0.003, 0.005, 0.007, 0.01\}$ at each
training step. We use the Lion optimizer~\cite{chen2024symbolic}
with constant learning rate $5 \times 10^{-6}$ and EMA decay
$0.9999$. The best checkpoint is selected by evaluating all saved
checkpoints (every 50K steps) on $10^5$ fresh samples at the target
error rate; the best checkpoints were typically found at
2.3--2.9M iterations depending on distance.
 
For experimental data from Sycamore, we follow a two-stage
protocol adapted from AlphaQubit~\cite{bausch2024learning,alphaqubit2}.
The first stage pretrains on synthetic data sampled
from a Detector Error Model (DEM) calibrated to the target
hardware via XEB calibration data,
using a round curriculum analogous to SI1000 (Stage~1:
$r \leq 9$, Stage~2: $r \leq 17$, Stage~3: $r \leq 25$, with
transitions at 150K and 300K iterations). The second stage
finetunes on real experimental data with a reduced learning rate
($2 \times 10^{-6}$) and zero weight decay.
 
Following AlphaQubit~2~\cite{alphaqubit2}, we apply BERT-style
input masking during SI1000 training and Sycamore pretraining:
with probability 0.8, we zero out 50\% of randomly selected defect
features. This regularization technique encourages the model to
learn robust representations from partial information.
 
For SI1000 and Sycamore, we scale the base learning rate according
to sequence length, following AlphaQubit~2~\cite{alphaqubit2}:
\begin{equation}
    \text{lr}_{\text{eff}} = \text{lr}_{\text{base}} \times
    0.8^{\log_2(N_s / 8)} \times 2^{\log_2(R / 24)},
\end{equation}
where $N_s$ is the number of stabilizers and $R$ is the number of
rounds, capped at a maximum scaling factor of $5\times$.

\section{Experiments}
\label{sec:experiments}

We evaluate the Sparse Mamba Decoder across four benchmarks of increasing 
complexity, from idealized noise models to real experimental data. 
Training was conducted on a mix of consumer and datacenter GPUs depending 
on model size: the depolarizing (${\sim}7.7$M parameters), uniform 
circuit-level (${\sim}7.6$M), and Sycamore (${\sim}7.6$M) models were 
trained on a single NVIDIA RTX 4090 (24\,GB), while the larger SI1000 
model (${\sim}16$M parameters for $d \geq 5$, with $R = 120$ rounds) 
required a single NVIDIA H200 NVL (141\,GB). Speed benchmarks under 
SI1000 noise (Section~\ref{sec:speed}) were measured on the H200; speed benchmarks under uniform circuit-level noise were measured on 
the RTX 4090. 

Architecture and training hyperparameters are summarized in 
Tables~\ref{tab:architecture} and~\ref{tab:training}, and 
Table~\ref{tab:decoder_comparison} situates SMD against existing 
ML-based decoders in terms of input representation, complexity, and 
hardware requirements. All circuit-level noise models and Monte Carlo 
sampling use the Stim simulator~\cite{gidney2021stim}; baseline 
matching decoders use PyMatching~\cite{higgott2023sparse}, Belief 
Matching~\cite{higgott2022beliefmatching}, and 
Tesseract~\cite{beni2025tesseract}. For uniform circuit-level 
(Section~\ref{sec:circuit_level_uniform}) and depolarizing 
(Section~\ref{sec:depolarizing}) benchmarks, we report the logical 
error rate (LER) as the per-shot failure fraction 
$\text{LER} = E_\text{fail}/N$, following Lange et 
al.~\cite{lange2023gnn}. For SI1000 
(Section~\ref{sec:circuit_level_si1000}) we report per-cycle LER 
following AlphaQubit~2~\cite{alphaqubit2}, and for Sycamore 
(Section~\ref{sec:sycamore}) per-round LER via log-fidelity 
regression following Lee et al.~\cite{lee2025scalable}. For SI1000 speed comparisons (Section~\ref{sec:speed}, 
Table~\ref{tab:speed} and Figure~\ref{fig:speed_accuracy}), we 
report the frame error rate (FER), identical to per-shot LER for a 
single memory experiment, following AlphaQubit~2~\cite{alphaqubit2}. 
The uniform circuit-level speed comparison 
(Figure~\ref{fig:speed_pareto_circuit}) uses per-shot LER as in 
Section~\ref{sec:circuit_level_uniform}. Unless otherwise stated, statistical 
uncertainties on Monte Carlo estimates are reported as 95\% binomial 
confidence intervals; for the Sycamore log-fidelity regression, 
uncertainty is estimated from the regression standard error.

\begin{table}[t]
\centering
\caption{Architecture hyperparameters across the four experimental
settings.}
\label{tab:architecture}
\small
\begin{tabular}{@{}lccccc@{}}
\toprule
\textbf{Parameter} & \textbf{Symbol} & \textbf{Sycamore} & \textbf{SI1000} & \textbf{Circuit-level} & \textbf{Depolarizing} \\
\midrule
Model dimension      & $\dmodel$  & 320  & 320/384$^*$  & 320  & 320 \\
Mamba layers         & $L$        & 4    & 4/6$^*$      & 4    & 4 \\
Mamba state dim      & $\dstate$  & 16   & 16   & 16   & 16 \\
Mamba conv width     & $\dconv$   & 4    & 4    & 4    & 4 \\
Mamba expand         & --         & 2    & 2    & 2    & 2 \\
Gate widening        & $\wgate$   & 5    & 5    & 5    & 5 \\
Input features       & --         & 13   & 13   & 13   & 13 \\
Readout type         & --         & ResBlock & ResBlock & MLP & Dual MLP \\
Output heads         & --         & 1    & 1    & 1    & 2 ($\lambda_Z$, $\lambda_X$) \\
Dropout              & --         & 0.1  & 0.1  & 0.1  & 0.1 \\
Parameters           & --         & ${\sim}7.6$M & 7.5/16M$^*$ & ${\sim}7.6$M & ${\sim}7.7$M \\
\bottomrule
\end{tabular}
\vspace{2pt}
\par\raggedright\footnotesize{$^*$SI1000: $\dmodel = 320$, $L = 4$, ${\sim}7.5$M params for $d = 3$; $\dmodel = 384$, $L = 6$, ${\sim}16$M params for $d \geq 5$.}
\end{table}

\begin{table}[t]
\centering
\caption{Training hyperparameters across the four experimental
settings. Sycamore uses DEM-based pretraining followed by
finetuning on experimental data; the iterations row reports the
DEM-pretraining stage. SI1000 training is stopped when validation 
LER plateaus, typically at 2.3--2.9M iterations depending on 
distance. The remaining settings train from scratch on synthetic data.}
\label{tab:training}
\small
\begin{tabular}{@{}lcccc@{}}
\toprule
\textbf{Parameter} & \textbf{Sycamore} & \textbf{SI1000} & \textbf{Circuit-level} & \textbf{Depolarizing} \\
\midrule
\multicolumn{5}{@{}l}{\textit{Training}} \\
Optimizer               & Lion    & Lion    & AdamW  & AdamW \\
Learning rate           & $5\!\times\!10^{-6}$ & $5\!\times\!10^{-6}$ & $5\!\times\!10^{-5}$--$10^{-4}$ & $1\!\times\!10^{-4}$ \\
LR schedule             & constant & constant & cosine & cosine \\
Iterations              & 1--4M$^\ddagger$ & 2.3--2.9M & --- & --- \\
Training epochs         & ---     & ---     & 600--1000  & 500 \\
Samples/epoch           & ---     & ---     & 1--6M  & 1--3M \\
Batch size              & 128--256 & 256--512 & 128--512 & 512 \\
Train error rates       & ---     & [.001--.01]  & [.001--.005] & [.01,.05,.10,.15] \\
Mixed rates             & ---     & yes     & yes    & yes \\
EMA decay               & 0.9999  & 0.9999  & off    & off \\
Weight decay            & $10^{-5}$ & $10^{-5}$ & $10^{-2}$ & $10^{-2}$ \\
Gradient clipping       & 1.0     & 1.0     & 1.0    & 1.0 \\
Curriculum stages       & 3 (150K/stage) & 3 (500K/stage) & --- & --- \\
Training noise model    & DEM     & SI1000  & uniform & depolarizing \\
\midrule
\multicolumn{5}{@{}l}{\textit{Finetuning (Sycamore only)}} \\
Finetune LR             & $2\!\times\!10^{-6}$ & --- & --- & --- \\
Finetune epochs         & 10      & ---     & ---    & --- \\
FT split ($d\!=\!3$)    & 50/50   & ---     & ---    & --- \\
FT split ($d\!=\!5$)    & 80/20   & ---     & ---    & --- \\
\midrule
\multicolumn{5}{@{}l}{\textit{Inference}} \\
Ensemble size           & 5/16$^*$ & 3$^\dagger$ & ---    & --- \\
\bottomrule
\end{tabular}
\vspace{2pt}
\par\raggedright\footnotesize{$^*$Sycamore: 5 models for $d=3$, 16 for $d=5$.\\
$^\dagger$SI1000: 3-checkpoint ensemble for $d=7$ only (steps 2.65M, 2.85M, 2.90M).\\
$^\ddagger$Sycamore: 1M iterations for $d=3$, 4M for $d=5$.}
\end{table}

\begin{table}[t]
\centering
\caption{Comparison of decoder architectures by input representation,
complexity, and hardware. Here $n$ denotes the syndrome size or number
of detection events, $B$ is the Tesseract beam-search width, and $M$
is the Libra ensemble size. Our sparse approach yields $\bigO(k)$
with $k \ll d^2 R$.}
\label{tab:decoder_comparison}
\small
\begin{tabular}{@{}lcccl@{}}
\toprule
\textbf{Decoder} & \textbf{Params} & \textbf{Input} & \textbf{Complexity} & \textbf{Hardware} \\
\midrule
\multicolumn{5}{@{}l}{\textit{Classical decoders (no learned parameters)}} \\
MWPM (PyMatching)        & ---    & Detection events  & $\bigO(n\log n)$  & CPU \\
Tesseract                & ---    & DEM graph         & $\bigO(n \cdot B)$ & CPU \\
Libra~\cite{jones2024libra} & ---  & Detection events & $\bigO(n\log n \cdot M)$ & CPU \\
\addlinespace
\multicolumn{5}{@{}l}{\textit{Neural decoders}} \\
AlphaQubit 1             & 5.4M   & Dense $d^2 \times R$ & $\bigO(d^4 R)$  & TPU v5e \\
AlphaQubit 2 (full)      & ${\sim}$32M & Dense $d^2 \times R$  & $\bigO(d^2 R)$  & Trillium TPU \\
AlphaQubit 2 (RT)        & ${\sim}$10M & Dense $d^2 \times R$ & $\bigO(d^2 R)$ & Trillium TPU \\
Dense Mamba              & ${\sim}$5--10M & Dense $d^2 \times R$ & $\bigO(d^2 R)$ & RTX 4090 \\
\addlinespace
\textbf{Proposed SMD} & $\mathbf{7.5\text{--}16}$\textbf{M} & \textbf{Sparse $k$ defects} & $\bm{\bigO(k)}$ & \textbf{RTX 4090 / H200} \\
\bottomrule
\end{tabular}
\end{table}

\subsection{Depolarizing Noise with Perfect Stabilizers}
\label{sec:depolarizing}

We first evaluate on the simplest noise model, following Lange et 
al.~\cite{lange2023gnn} (Figure~8 therein). Data qubits experience 
depolarizing noise $\mathcal{E}(\rho) = (1-p)\rho + 
\frac{p}{3}(X\rho X + Y\rho Y + Z\rho Z)$ with perfect (noiseless) 
stabilizer measurements, producing a purely spatial decoding problem 
with a single round. The decoder predicts both logical observables 
$\lambda_Z$ and $\lambda_X$ using the dual-head variant. For this 
single-round setting, the 13th feature (reconstructed measurement) 
equals 1.0 at all defect positions, since a single detection event 
that fired trivially gives $m_{i,1} = d_{i,1} = 1$.

\begin{figure}[ht!]
    \centering
    \includegraphics[width=\linewidth]{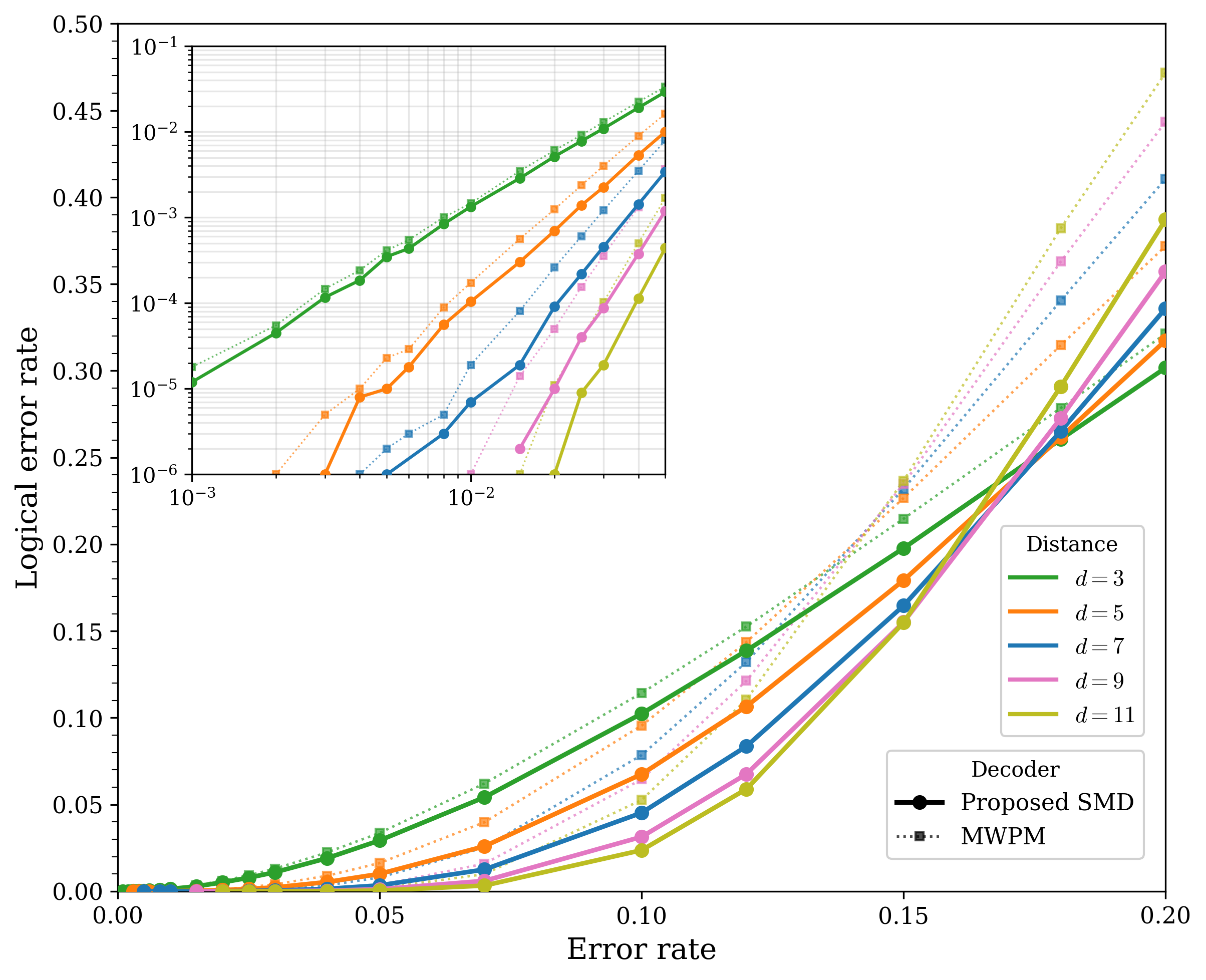}
    \caption{Logical error rate under depolarizing noise with perfect 
    stabilizer measurements. The Sparse Mamba Decoder (solid lines) 
    with dual prediction heads for $\lambda_Z$ and $\lambda_X$ is 
    compared against MWPM (dotted) at distances $d \in \{3, 5, 7, 9, 11\}$. 
    Each data point is evaluated over $10^6$--$3 \times 10^6$ fresh samples. 
    \emph{Inset:} Log-log view of the low-$p$ region, showing the widening 
    gap between SMD and MWPM with increasing distance.}
    \label{fig:depolarizing}
\end{figure}

Figure~\ref{fig:depolarizing} shows the logical error rates across all 
distances and error rates. SMD outperforms MWPM at every distance and 
error rate tested, with improvement increasing monotonically with code 
distance: from 6--15\% at $d = 3$ to 17--91\% at $d = 11$.

A striking finding is that MWPM, the Tesseract near-MLD 
decoder~\cite{beni2025tesseract}, and Belief 
Matching~\cite{higgott2022beliefmatching} all achieve nearly identical 
logical error rates across all distances and error rates 
(Table~\ref{tab:depolarizing_results}). This suggests that MWPM is 
already near-optimal among factorized independent-basis decoders for 
the rotated surface code with depolarizing noise and perfect stabilizers, 
consistent with the observation by Lange et al.\ that their GNN decoder 
converges to approximate MLD accuracy at $d \leq 5$.

The classical decoders decode $\lambda_Z$ and $\lambda_X$ independently: 
MWPM performs separate matching on the Z-stabilizer and X-stabilizer 
syndrome graphs. Our decoder, in contrast, receives both syndrome types 
as input and predicts $(\lambda_Z, \lambda_X)$ jointly. Under depolarizing 
noise, $Y$ errors (occurring with probability $p/3$) simultaneously 
trigger both Z-type and X-type stabilizers, creating correlated defect 
patterns across the two syndrome types. The Mamba backbone naturally 
captures these $Y$-error correlations through its joint sequence 
processing, which independent matching decoders structurally cannot 
exploit. This explains how SMD surpasses the near-optimal accuracy of 
independent decoding by 40--91\% at larger distances. We emphasize that 
joint decoding is a structural advantage of neural decoders: 
matching-based decoders (MWPM, Tesseract, Belief Matching) are 
fundamentally factorized into independent $X$- and $Z$-graph matchings, 
whereas a single neural backbone can model the cross-basis correlations 
introduced by $Y$ errors. The 40--91\% improvement therefore reflects 
both the sparse representation and the architectural capacity to exploit 
correlations inaccessible to factorized matching decoders.

\begin{table}[t]
\centering
\caption{Decoder comparison under depolarizing noise with perfect 
stabilizers (combined failure: wrong $\lambda_Z$ or wrong $\lambda_X$). 
All three classical decoders achieve nearly identical error rates, 
confirming MWPM is near-optimal for independent-basis decoding. SMD 
outperforms all through joint $(\lambda_Z, \lambda_X)$ prediction. 
Bold indicates the best result per cell. A dash indicates fewer than 
one failure in $3 \times 10^6$ samples.}
\label{tab:depolarizing_results}
\small
\setlength{\tabcolsep}{3pt}
\begin{tabular}{@{}llccccc@{}}
\toprule
& & $d\!=\!3$ & $d\!=\!5$ & $d\!=\!7$ & $d\!=\!9$ & $d\!=\!11$ \\
\midrule
\multirow{4}{*}{\rotatebox{90}{$p\!=\!.01$}}
& MWPM            & 1.5e-3 & 1.7e-4 & 1.9e-5 & 1e-6   & --- \\
& Tesseract       & 1.5e-3 & 1.7e-4 & 1.9e-5 & 1e-6   & --- \\
& Belief Match.   & 1.5e-3 & 1.8e-4 & 2.0e-5 & 1e-6   & --- \\
& \textbf{Mamba}  & \textbf{1.3e-3} & \textbf{1.0e-4} & \textbf{7e-6} & ---    & --- \\
\midrule
\multirow{4}{*}{\rotatebox{90}{$p\!=\!.03$}}
& MWPM            & 1.3e-2 & 4.0e-3 & 1.2e-3 & 3.6e-4 & 1.0e-4 \\
& Tesseract       & 1.3e-2 & 4.1e-3 & 1.2e-3 & 3.7e-4 & 1.0e-4 \\
& Belief Match.   & 1.3e-2 & 4.1e-3 & 1.2e-3 & 3.6e-4 & 1.1e-4 \\
& \textbf{Mamba}  & \textbf{1.1e-2} & \textbf{2.3e-3} & \textbf{4.6e-4} & \textbf{8.8e-5} & \textbf{1.9e-5} \\
\midrule
\multirow{4}{*}{\rotatebox{90}{$p\!=\!.05$}}
& MWPM            & 3.4e-2 & 1.6e-2 & 8.0e-3 & 3.7e-3 & 1.7e-3 \\
& Tesseract       & 3.5e-2 & 1.7e-2 & 7.9e-3 & 3.8e-3 & 1.7e-3 \\
& Belief Match.   & 3.5e-2 & 1.7e-2 & 8.1e-3 & 3.7e-3 & 1.9e-3 \\
& \textbf{Mamba}  & \textbf{2.9e-2} & \textbf{1.0e-2} & \textbf{3.4e-3} & \textbf{1.2e-3} & \textbf{4.4e-4} \\
\midrule
\multirow{4}{*}{\rotatebox{90}{$p\!=\!.10$}}
& MWPM            & 1.1e-1 & 9.6e-2 & 7.9e-2 & 6.5e-2 & 5.3e-2 \\
& Tesseract       & 1.2e-1 & 9.8e-2 & 8.0e-2 & 6.6e-2 & 5.3e-2 \\
& Belief Match.   & 1.2e-1 & 1.0e-1 & 8.1e-2 & 6.6e-2 & 5.4e-2 \\
& \textbf{Mamba}  & \textbf{1.0e-1} & \textbf{6.8e-2} & \textbf{4.5e-2} & \textbf{3.1e-2} & \textbf{2.4e-2} \\
\bottomrule
\end{tabular}
\end{table}

\subsection{Circuit-Level Noise (Uniform)}
\label{sec:circuit_level_uniform}

We next evaluate on circuit-level noise following Lange et 
al.~\cite{lange2023gnn} (Figure~4 therein), where all operations 
(two-qubit gates, single-qubit gates, measurements, resets) experience 
the same depolarizing rate $p$. The number of stabilizer measurement 
rounds equals the code distance ($R = d$), and the decoder predicts a 
single logical observable $\lambda_Z$ from a Z-basis memory experiment.

We compare against MWPM (PyMatching~\cite{higgott2022pymatching}), the 
standard matching-based baseline. SMD uses a single-head variant 
predicting $\lambda_Z$ only, with $\dmodel = 320$, $L = 4$ layers, and 
expand factor 2. Training uses AdamW with cosine learning rate annealing 
over 600--1000 epochs, with $1$--$6 \times 10^6$ freshly generated 
samples per epoch and error rates sampled uniformly from $p \in \{0.001, 
0.002, 0.003, 0.004, 0.005\}$. Each data point is evaluated over 
$4$--$6 \times 10^6$ fresh samples.

\begin{figure}[ht!]
    \centering
    \includegraphics[width=\linewidth]{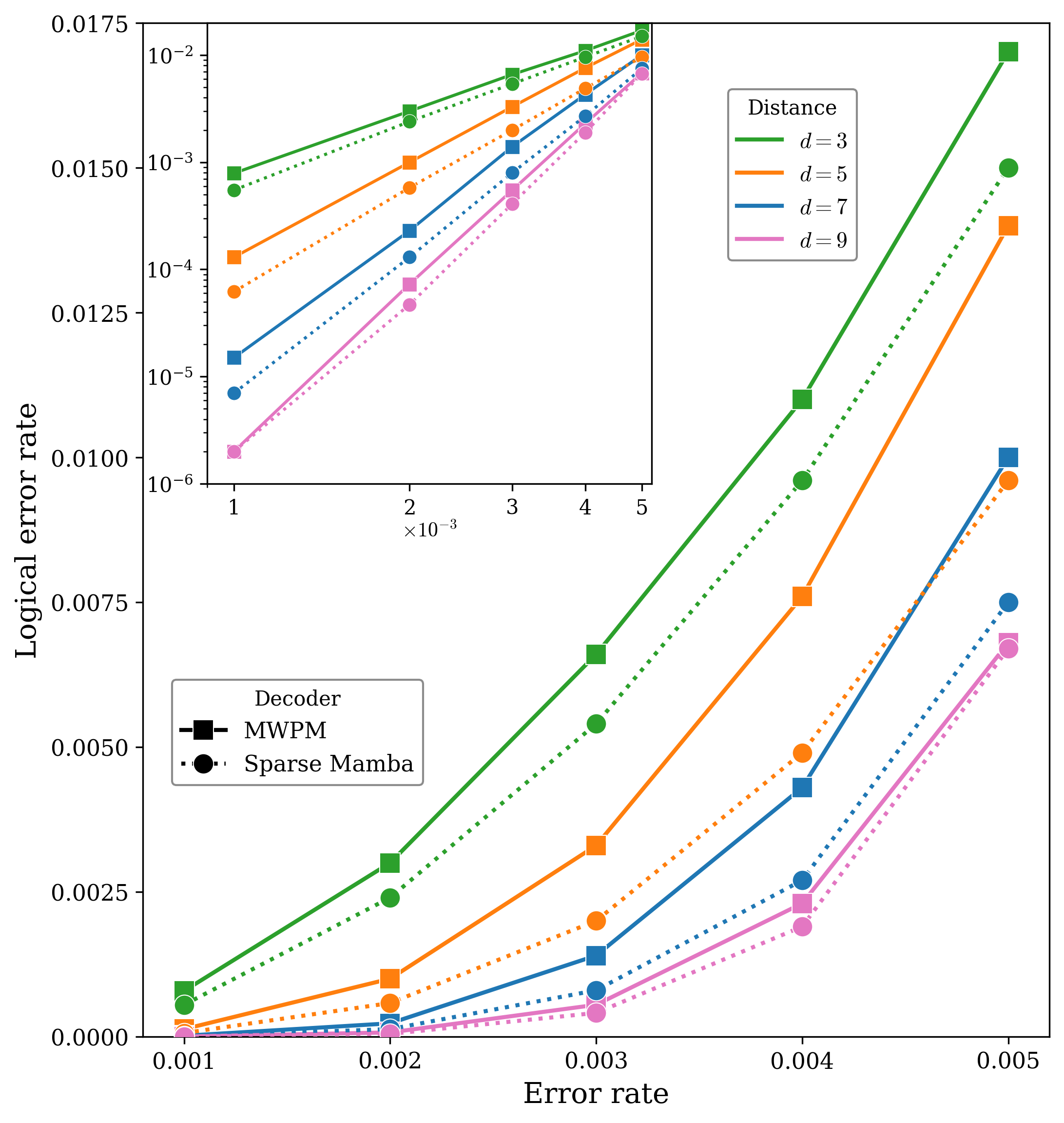}
    \caption{Logical error rate under uniform circuit-level noise with 
    $R = d$ stabilizer measurement rounds. SMD (dotted circles) vs 
    MWPM (solid squares) at distances $d \in \{3, 5, 7, 9\}$ and 
    error rates $p \in [0.001, 0.005]$. SMD outperforms MWPM at all 
    distances and rates.}
    \label{fig:circuit_uniform}
\end{figure}

SMD outperforms MWPM at every distance and error rate tested 
(Figure~\ref{fig:circuit_uniform}), with LER ratios (SMD/MWPM) 
ranging from 0.70 at $d = 3$ to 0.47 at $d = 7$ at the lowest 
error rate ($p = 0.001$). The improvement is consistent across all 
rates: at $p = 0.003$, SMD reduces the logical error rate by 18\% 
at $d = 3$, 39\% at $d = 5$, 44\% at $d = 7$, and 25\% at $d = 9$. 
At $d = 9$, $p = 0.005$, SMD essentially matches MWPM (ratio 0.98). 
A speed--accuracy comparison with Belief 
Matching~\cite{higgott2022beliefmatching}, which achieves lower 
rates at $d \geq 7$ but at 18--230$\times$ higher latency for 
$d \geq 5$, is presented in Section~\ref{sec:speed}.

The GNN decoder of Lange et al.~\cite{lange2023gnn} trained each model 
for 1000 epochs with $10^7$ freshly generated samples per epoch, 
totaling $10^{10}$ training samples per distance. In contrast, SMD 
uses $6 \times 10^8$ samples at $d = 3$ (600 epochs $\times$ $10^6$), 
$1.2 \times 10^9$ at $d = 5$ (600 $\times$ $2 \times 10^6$), 
$1.8 \times 10^9$ at $d = 7$ (600 $\times$ $3 \times 10^6$), and 
$6 \times 10^9$ at $d = 9$ (1000 $\times$ $6 \times 10^6$)---a 
1.7--17$\times$ reduction in training data. Despite this, SMD 
outperforms MWPM at all distances and error rates, whereas the GNN 
decoder of Lange et al.\ reports performance approximately matching 
MWPM (their Figure~4 does not provide exact numerical values, 
precluding a direct quantitative comparison). At $d = 9$, where our 
training budget is 60\% of the reference, SMD outperforms MWPM for 
$p \leq 0.004$ (ratios 0.63--0.86) and essentially matches MWPM at 
$p = 0.005$ (ratio 0.98); the remaining gap is consistent with the 
reduced training budget rather than a fundamental architectural 
limitation.

\subsection{Circuit-Level Noise (SI1000)}
\label{sec:circuit_level_si1000}

To enable direct comparison with AlphaQubit~2~\cite{alphaqubit2} and 
the Tesseract near-MLD baseline~\cite{beni2025tesseract} (compared 
on speed--accuracy in Section~\ref{sec:speed}), we evaluate on the 
SI1000 noise model (a superconducting-inspired noise model with 
asymmetric error rates, Table~\ref{tab:si1000}), which is the 
standard benchmark used by recent high-accuracy decoders. We test at $p = 1.5 \times 10^{-3}$, representative of near-term hardware operating conditions and well below 
the SI1000 surface-code threshold ($p_{\mathrm{th}} \approx 5 \times 10^{-3}$ 
for our $R = 120$ setup), with code distances $d \in \{3, 5, 7\}$, 
$R = 120$ stabilizer measurement rounds, and $5 \times 10^5$ Monte Carlo 
shots per configuration.

\begin{table}[h]
\centering
\caption{SI1000 noise model error rates relative to the base two-qubit 
gate rate $p$ (following~\cite{alphaqubit2}).}
\label{tab:si1000}
\small
\begin{tabular}{@{}lcc@{}}
\toprule
\textbf{Operation} & \textbf{Rate} & \textbf{Description} \\
\midrule
Two-qubit gate (CZ) & $p$     & Two-qubit depolarizing \\
Single-qubit gate    & $p/10$  & Single-qubit depolarizing \\
Measurement readout  & $5p$    & Bit-flip before readout \\
Reset                & $2p$    & Bit-flip after reset \\
Idle (per round)     & $4.2p$  & Combined idle depolarizing \\
\bottomrule
\end{tabular}
\end{table}

For the SI1000 benchmark, we train with a three-stage curriculum
over measurement rounds: the maximum round count increases from
$r \leq 9$ (Stage~1) to $r \leq 25$ (Stage~2) to $r \leq 120$
(Stage~3) at transitions of 500K and 1M iterations. We use the 
Lion optimizer~\cite{chen2024symbolic} with constant learning rate 
$5 \times 10^{-6}$ and EMA decay $0.9999$. The best checkpoint is 
selected by evaluating all saved checkpoints (every 50K steps) on 
$10^5$ fresh samples at the target error rate; training is stopped 
when validation LER plateaus, with the best checkpoints typically 
found at 2.3--2.9M iterations depending on distance.

For $d = 7$, where single-model performance is close to MWPM, we employ 
multi-checkpoint ensembling: we average the output logits of three models 
taken from different training steps (2.65M, 2.85M, 2.90M), which provide 
genuine diversity since models at different points in training have 
learned qualitatively different feature representations. This approach 
is analogous to the multi-seed ensembling used for Sycamore 
(Section~\ref{sec:sycamore}), but exploits checkpoint diversity rather 
than seed diversity.

Table~\ref{tab:si1000_results} presents the SI1000 results. At $d = 3$ 
and $d = 5$, the single-model SMD outperforms MWPM, 
achieving LER ratios of $0.65$ and $0.51$ respectively. At $d = 7$, the 
single model approximately matches MWPM (ratio $0.99$); the 
three-checkpoint ensemble reduces the LER to $1.44 \times 10^{-4}$, 
a $16\%$ improvement over MWPM though within statistical uncertainty 
($\pm 24\%$ relative 95\% CI at this LER and sample size).
Correlated Matching (Belief Matching) provides 
a stronger baseline at $d = 7$, achieving $7.45 \times 10^{-5}$ (ratio 
$0.44$), but at $d \leq 5$ SMD outperforms it: ratio $0.65$ vs.\ $0.71$ 
at $d = 3$ and $0.51$ vs.\ $0.55$ at $d = 5$. Ensembling provides 
meaningful gains only when the single model is uncertain or near 
saturation; we additionally evaluated multi-checkpoint ensembling at 
$d = 3$ and $d = 5$ and observed no significant further improvement, 
consistent with the single-model SMD already being substantially 
better than MWPM at these distances.

We also compare against a Libra-style ensemble 
decoder~\cite{jones2024libra}, which creates diversity by perturbing the 
edge weights of the detector error model and combining multiple MWPM 
solutions via majority vote. Despite using 7 ensemble members, the 
Libra-style decoder shows negligible improvement over standard MWPM at 
all distances ($< 0.5\%$), confirming that the matching-based approach 
has saturated its accuracy on this noise model. In contrast, our neural 
ensemble captures error correlations inaccessible to matching-based 
methods, yielding a $16\%$ LER reduction at $d = 7$.

\begin{table}[t]
\centering
\caption{Logical error rate per cycle under SI1000 noise at
$p = 1.5 \times 10^{-3}$, $R = 120$. Ratio is LER relative to MWPM.
Relative 95\% binomial CIs are approximately $\pm 7\%$, $\pm 16\%$,
and $\pm 23\%$ at $d = 3, 5, 7$ respectively. \textbf{Bold} indicates
best neural decoder per distance; classical matching baselines shown
for reference.}
\label{tab:si1000_results}
\small
\begin{tabular}{@{}lccc@{}}
\toprule
\textbf{Decoder} & $d=3$ & $d=5$ & $d=7$ \\
\midrule
MWPM (PyMatching) 
    & $2.21 \times 10^{-3}$ 
    & $6.20 \times 10^{-4}$ 
    & $1.71 \times 10^{-4}$ \\
Corr.\ Matching$^\dagger$
    & $1.57 \times 10^{-3}$ \scriptsize{(0.71$\times$)}
    & $3.42 \times 10^{-4}$ \scriptsize{(0.55$\times$)}
    & $7.45 \times 10^{-5}$ \scriptsize{(0.44$\times$)} \\
Libra-7 (MWPM ensemble) 
    & $2.21 \times 10^{-3}$ \scriptsize{(1.00$\times$)}
    & $6.20 \times 10^{-4}$ \scriptsize{(1.00$\times$)}
    & $1.71 \times 10^{-4}$ \scriptsize{(1.00$\times$)} \\
\addlinespace
Proposed SMD (single) 
    & $\mathbf{1.43 \times 10^{-3}}$ \scriptsize{(0.65$\times$)}
    & $\mathbf{3.14 \times 10^{-4}}$ \scriptsize{(0.51$\times$)}
    & $1.70 \times 10^{-4}$ \scriptsize{(0.99$\times$)} \\
Proposed SMD (ensemble) 
    & $1.43 \times 10^{-3}$ \scriptsize{(0.65$\times$)}
    & $3.16 \times 10^{-4}$ \scriptsize{(0.51$\times$)}
    & $\mathbf{1.44 \times 10^{-4}}$ \scriptsize{(0.84$\times$)} \\
\bottomrule
\multicolumn{4}{l}{\scriptsize $^\dagger$Evaluated with $N = 10^4$ due to cost (327\,ms/shot).}
\end{tabular}
\end{table}

\subsubsection*{Round dependence}

A distinctive feature of the sparse decoder is that its performance
depends not only on code distance but also on the number of measurement
rounds, which directly controls the defect sequence length $k$.
Table~\ref{tab:round_dependence} shows the LER ratio (SMD/MWPM) as a
function of rounds for $d = 5$ and $d = 7$.

At $d = 5$, SMD \emph{loses} to MWPM at short rounds ($r \leq 25$,
ratio $> 1$) but \emph{wins} decisively at long rounds ($r \geq 50$,
ratio $\approx 0.5$). This is because at short rounds with low $p$, the
expected number of defects is very small ($k \approx 7$ at $r = 10$),
providing insufficient signal for the learned model---MWPM's structural
knowledge of the error model compensates for its inability to learn
correlations. At $d = 7$, the sparse decoder beats MWPM at $r \leq 80$
(ratio $0.77$--$0.84$) and matches MWPM at $r = 120$
(ratio $0.99$). Only beyond the training range at $r = 140$ does
performance degrade significantly (ratio $1.40$), as the defect count
exceeds the model's training distribution. Multi-checkpoint ensembling recovers strong performance even at $r = 120$, reducing the ratio to $0.84$. Notably, at $d = 5$
the model generalizes beyond its training range: performance at
$r = 140$ (not seen during training, which used $r \leq 120$) remains
strong (ratio $0.54$), demonstrating that the learned representations
extrapolate to longer sequences.

Table~\ref{tab:error_rate_dependence} shows the LER ratio as a function
of physical error rate. At $d = 3$ and $d = 5$, SMD outperforms MWPM
across all sub-threshold rates, with low ratios of
$0.65$--$0.71$ at $d = 3$ and $0.51$--$0.61$ at $d = 5$. At $d = 7$,
SMD wins only at the lowest rate ($p = 0.001$, ratio $0.90$) where the
syndrome is sparsest ($k \approx 185$ mean defects). At higher rates,
the growing defect count exceeds the model's effective capacity,
consistent with the round-dependence analysis above.

\begin{table}[t]
\centering
\caption{LER ratio (SMD / MWPM) as a function of measurement rounds
under SI1000 noise at $p = 1.5 \times 10^{-3}$, $N = 10^5$ shots.
Values below 1.0 indicate SMD outperforms MWPM. Model trained with
$r \leq 120$; $r = 140$ tests out-of-distribution generalization.}
\label{tab:round_dependence}
\small
\begin{tabular}{@{}lcccccc@{}}
\toprule
& $r=10$ & $r=25$ & $r=50$ & $r=80$ & $r=120$ & $r=140^*$ \\
\midrule
$d = 5$ & 7.93 & 1.88 & \textbf{0.72} & \textbf{0.52} & \textbf{0.52} & \textbf{0.54} \\
$d = 7$ & \textbf{0.81} & \textbf{0.84} & \textbf{0.77} & \textbf{0.84} & 0.99 & 1.40 \\
\bottomrule
\multicolumn{7}{l}{\scriptsize $^*$Beyond training range ($r_{\max} = 120$).}
\end{tabular}
\end{table}

\begin{table}[t]
\centering
\caption{LER ratio (SMD / MWPM) as a function of physical error rate
under SI1000 noise at $R = 120$, $N = 10^5$ shots ($N = 5 \times 10^5$
for $d = 7$). Values below 1.0 indicate SMD outperforms MWPM.}
\label{tab:error_rate_dependence}
\small
\begin{tabular}{@{}lcccc@{}}
\toprule
& $p=0.001$ & $p=0.0015$ & $p=0.002$ & $p=0.003$ \\
\midrule
$d = 3$ & \textbf{0.65} & \textbf{0.66} & \textbf{0.68} & \textbf{0.71} \\
$d = 5$ & \textbf{0.51} & \textbf{0.51} & \textbf{0.53} & \textbf{0.61} \\
$d = 7$ & \textbf{0.90} & 0.99 & 1.45 & 2.08 \\
\bottomrule
\end{tabular}
\end{table}

\subsection{Sycamore Experimental Data}
\label{sec:sycamore}

We evaluate on the publicly available Sycamore memory experiment 
dataset~\cite{google2023suppressing,bausch2024learning}, which comprises 
50,000 shots per round setting across four distance-3 patches 
(north/south/east/west) and one distance-5 patch, for both X- and 
Z-basis memory experiments with rounds $R \in \{1, 3, 5, \ldots, 25\}$. 
This experiment provides a direct comparison with the dense Mamba 
decoder of Lee et al.~\cite{lee2025scalable}, which reports LERs 
of $2.98 \times 10^{-2}$ (distance~3) and $3.03 \times 10^{-2}$ 
(distance~5), averaged across X and Z bases. We follow an identical 
evaluation protocol: DEM-based pretraining from XEB calibration data, 
finetuning on experimental data, and LER computation via log-fidelity 
regression across rounds $\{3, 5, \ldots, 25\}$, excluding round~1 due 
to time-boundary effects.

We pretrain using a curriculum learning strategy over rounds: 
Stage~1 ($r \in \{3, 5, 7, 9\}$), Stage~2 ($r \leq 17$), and 
Stage~3 ($r \leq 25$), with transitions at 150K and 300K iterations 
following Lee et al.~\cite{lee2025scalable}. For distance~3, we 
pretrain for 1M iterations with batch size 256; for distance~5, we 
extend pretraining to 4M iterations with batch size 128, as the single 
distance-5 patch provides only 600K experimental samples (compared to 
2.4M for the four distance-3 patches), requiring stronger pretrained 
representations to compensate for limited finetuning data. During 
finetuning, we use the Lion optimizer with learning rate 
$2 \times 10^{-6}$ and zero weight decay. For distance~3, we use a 
50/50 train/evaluation split (1.2M training samples); for distance~5, 
we use an 80/20 split (480K training samples) to maximize the training 
set given the limited data from a single patch.

Following standard practice in neural QEC 
decoding~\cite{bausch2024learning,alphaqubit2}, we improve prediction 
accuracy through model ensembling. We finetune multiple models from the 
same pretrained checkpoint using different random seeds (which produce 
different train/evaluation splits), then average their output logits at 
inference time. Additionally, we finetune models from pretrained 
checkpoints at different training stages (3M, 3.5M, and 4M steps for 
distance~5), which provides genuine model diversity beyond what seed 
variation alone can achieve---models at different points in pretraining 
have learned qualitatively different feature representations. Our final 
ensemble combines 16 models per basis (10 from the 4M checkpoint with 
different seeds, plus 3 each from the 3M and 3.5M checkpoints). For 
distance~3, we ensemble 5 models per basis.

\begin{figure}[ht!]
    \centering
    \includegraphics[width=\linewidth]{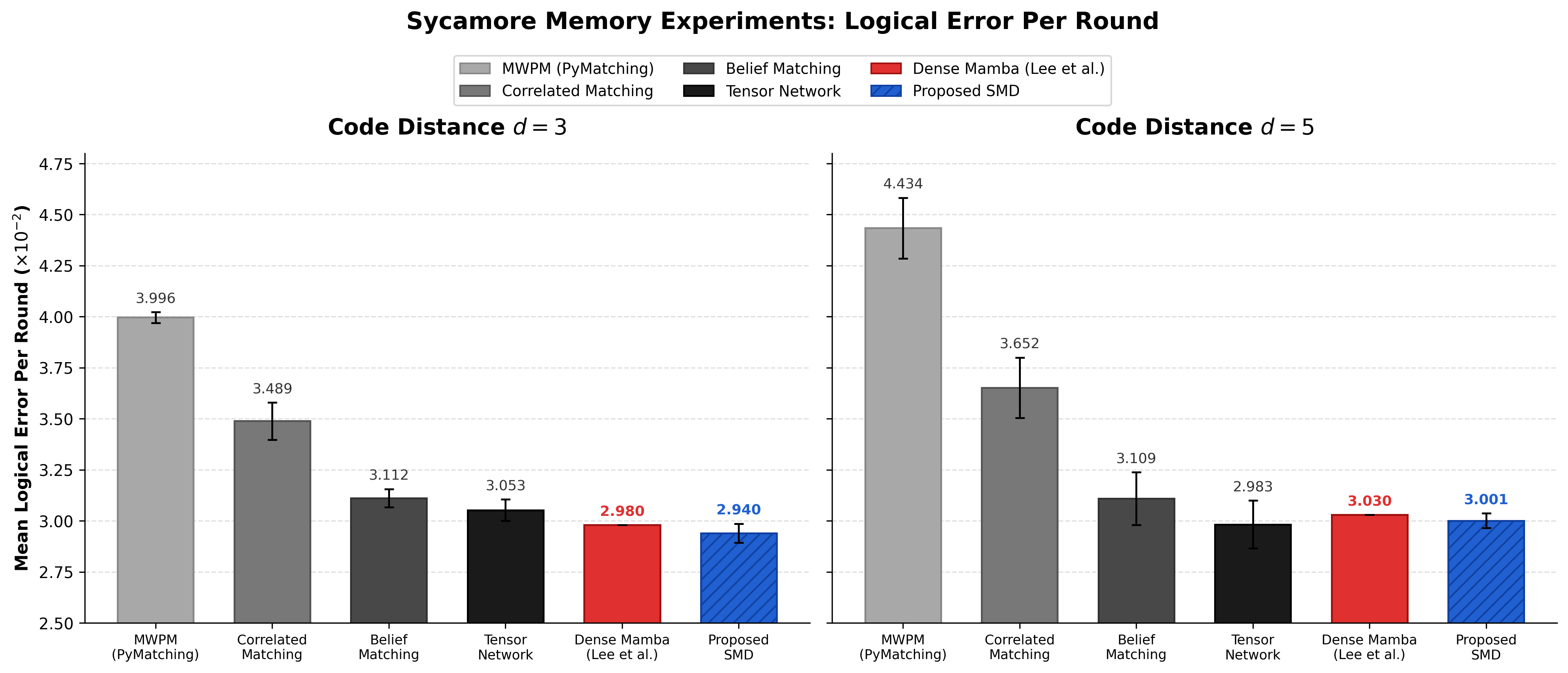}
    \caption{Mean logical error per round on the Google Sycamore 
    experimental dataset at code distances 3 and~5. Results are averaged 
    across X and Z bases. Error bars indicate the spread between bases. 
    The proposed Sparse Mamba Decoder (blue, hatched) matches or 
    outperforms the dense Mamba decoder of Lee et 
    al.~\cite{lee2025scalable} (red) at both distances, despite 
    processing only active defects with $\bigO(k)$ complexity compared 
    to $\bigO(d^2 R)$ for dense approaches.}
    \label{fig:sycamore}
\end{figure}

The proposed SMD achieves a mean LER of $2.940 \times 10^{-2}$ at 
distance~3 and $3.001 \times 10^{-2}$ at distance~5 (ensemble, averaged 
across bases), matching or surpassing the dense Mamba decoder of 
Lee et al.\ ($2.98 \times 10^{-2}$ and $3.03 \times 10^{-2}$, 
respectively) despite using a sparse input representation without 
dilated 2D convolutions. At distance~3, our decoder outperforms all 
baselines including the tensor network decoder ($3.053 \times 10^{-2}$). 
At distance~5, our ensemble result surpasses belief matching 
($3.109 \times 10^{-2}$) and the dense Mamba decoder, and comes within 
$0.6\%$ of the tensor network decoder ($2.983 \times 10^{-2}$). 
Single-model results without ensembling also demonstrate strong 
performance: $3.015 \times 10^{-2}$ at distance~3 and 
$3.130 \times 10^{-2}$ at distance~5. The single model surpasses 
correlated matching and belief matching at distance~3 and 
correlated matching at distance~5; ensembling is required to 
surpass belief matching and dense Mamba at distance~5. Ensembling provides a ${\sim}2.5\%$ relative LER improvement at 
distance~3 and ${\sim}4\%$ at distance~5. SMD ensemble achieves 
the lowest LER among all compared decoders at distance~3, and 
the second-lowest at distance~5 (behind only the tensor network).
Figure~\ref{fig:sycamore} summarizes the comparison.

\subsection{Computational Efficiency}
\label{sec:speed}

\begin{figure}[ht!]
    \centering
    \includegraphics[width=0.85\linewidth]{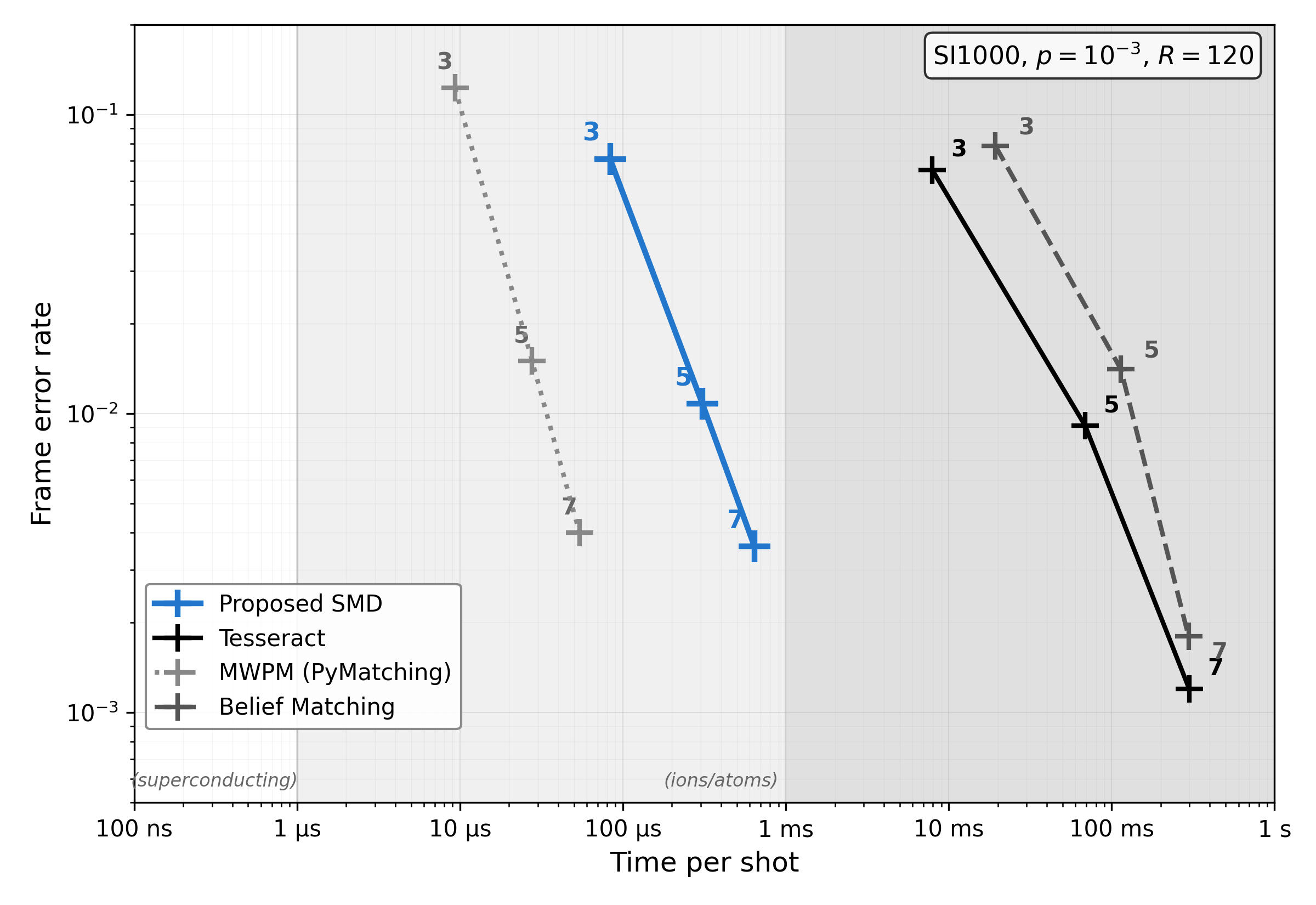}
    \caption{Speed--accuracy Pareto front for MWPM, Belief Matching, 
    Tesseract, and SMD at $d \in \{3, 5, 7\}$ under SI1000 noise at 
    $p = 10^{-3}$, measured on NVIDIA H200 NVL. Lower-left is better. 
    Vertical lines at $1\,\mu$s and $1$\,ms separate the 
    superconducting, trapped-ion/atom, and offline regimes. SMD 
    occupies the ions/atoms regime (100\,$\mu$s--1\,ms), achieving 
    lower FER than MWPM while running 95--467$\times$ faster than 
    Tesseract and 232--463$\times$ faster than Belief Matching. Distance labels indicate 
    $d \in \{3, 5, 7\}$.}
    \label{fig:speed_accuracy}
\end{figure}

A key advantage of the sparse representation is reduced computational 
cost at physically relevant error rates. Table~\ref{tab:speed} reports 
end-to-end decoding time and FER for MWPM, Belief 
Matching~\cite{higgott2022beliefmatching}, Tesseract, and SMD at 
$d \in \{3, 5, 7\}$ under SI1000 noise at $p = 10^{-3}$, $R = 120$, on 
a single NVIDIA H200 NVL GPU. SMD achieves two-to-three orders of magnitude speedup over near-MLD decoders while approaching their accuracy 
(Figure~\ref{fig:speed_accuracy}).

\begin{table}[t]
\centering
\caption{End-to-end decoding time per shot and frame error rate (FER)
under SI1000 noise at $p = 10^{-3}$, $R = 120$. MWPM, BM, and
Tesseract on a single CPU core; SMD on H200 NVL GPU including
feature-extraction time. See Table~\ref{tab:si1000_results} for
definitive accuracy at $N = 5 \times 10^5$.}
\label{tab:speed}
\small
\setlength{\tabcolsep}{4pt}
\begin{tabular}{@{}lcccccc@{}}
\toprule
 & \multicolumn{2}{c}{$d=3$} & \multicolumn{2}{c}{$d=5$} & \multicolumn{2}{c}{$d=7$} \\
\cmidrule(lr){2-3}\cmidrule(lr){4-5}\cmidrule(lr){6-7}
Decoder & time/shot & FER & time/shot & FER & time/shot & FER \\
\midrule
MWPM                & 9.3\,$\mu$s  & 12.3\% & 27.7\,$\mu$s & 1.50\% & 54.1\,$\mu$s & 0.40\% \\
Belief Matching     & 19.4\,ms     & 7.9\%  & 113.8\,ms    & 1.4\%  & 296.5\,ms    & 0.18\% \\
Tesseract           & 7.94\,ms     & 6.5\%  & 68.9\,ms     & 0.91\% & 299.3\,ms    & 0.12\% \\
\textbf{Proposed SMD} (b=1024) & \textbf{83\,$\mu$s} & \textbf{7.1\%} & \textbf{307\,$\mu$s} & \textbf{1.08\%} & \textbf{641\,$\mu$s} & \textbf{0.36\%} \\
\midrule
Speedup (Tesseract / SMD) & 95$\times$  & --- & 224$\times$ & --- & 467$\times$ & --- \\
Speedup (BM / SMD)        & 232$\times$ & --- & 370$\times$ & --- & 463$\times$ & --- \\
\bottomrule
\end{tabular}
\end{table}

SMD's defect-centric representation yields favorable scaling with code
distance. The SMD-to-Tesseract speedup grows from 95$\times$ at $d = 3$
to 467$\times$ at $d = 7$, while SMD-to-BM grows from 232$\times$ to
463$\times$. Notably, SMD captures a substantial fraction of Tesseract's accuracy 
advantage at orders of magnitude lower latency: at $d = 3$, SMD closes 
roughly $90\%$ of the FER gap between MWPM and Tesseract ($12.3\% \to 
7.1\%$ vs.\ $6.5\%$); at $d = 5$, $\sim 71\%$ of the gap is closed at 
$224\times$ lower latency than Tesseract. This follows directly from algorithmic complexity: 
Tesseract's beam search cost grows with the detector graph size 
($\Theta(d^2 R)$) and error density, whereas SMD's forward pass scales 
with the number of active defects $k$, which grows as 
$\bigO(p \cdot d^2 R)$. At fixed $p$, the ratio $k / (d^2 R)$ remains 
approximately constant across distances (Table~\ref{tab:sparsity}), so 
SMD's per-shot cost grows much more slowly with $d$ than competing 
decoders.  

\begin{figure}[ht!]
    \centering
    \includegraphics[width=0.85\linewidth]{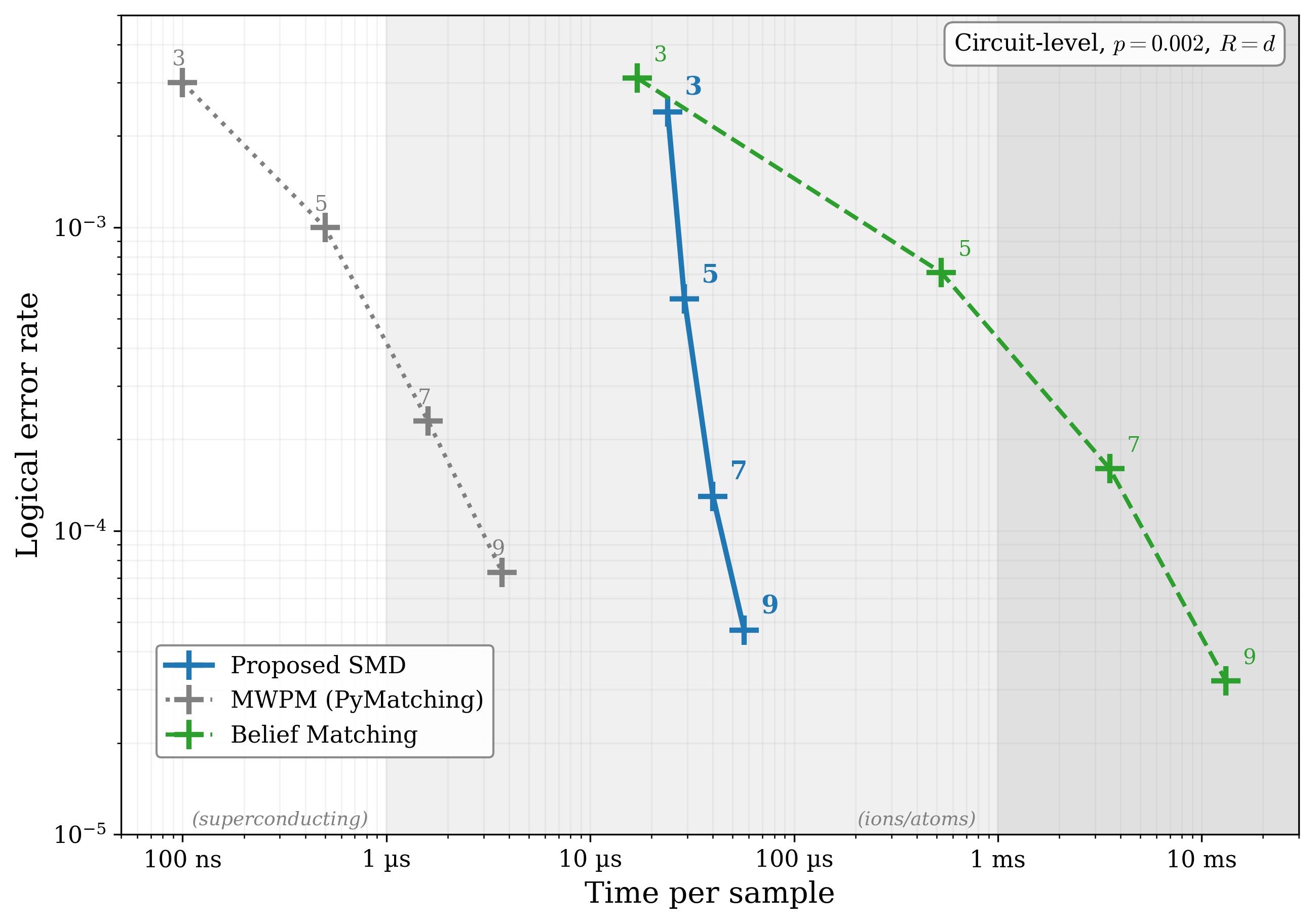}
    \caption{Speed--accuracy Pareto front under uniform circuit-level 
    noise at $p = 0.002$, $R = d$. SMD on RTX 4090; MWPM and BM on 
    a single CPU core. Lower-left is better. SMD (blue) is 
    18--230$\times$ faster than Belief Matching (green) for $d \geq 5$ 
    while outperforming MWPM (gray) in accuracy at all distances.}
    \label{fig:speed_pareto_circuit}
\end{figure}

Under uniform circuit-level noise at $p = 0.002$ 
(Figure~\ref{fig:circuit_uniform}), SMD latency increases only 
2.4$\times$ from $d = 3$ to $d = 9$ (24\,$\mu$s to 57\,$\mu$s on an 
NVIDIA RTX 4090), compared to 773$\times$ for Belief Matching 
(17\,$\mu$s to 13.1\,ms on a single CPU core). At $d = 9$, SMD 
decodes 230$\times$ faster than Belief Matching, while outperforming 
both MWPM and Belief Matching in accuracy at $d \leq 7$ 
(Figure~\ref{fig:speed_pareto_circuit}).

At single-shot latency (batch size 1), SMD requires 4--12\,ms per shot, 
dominated by GPU kernel-launch overhead rather than compute. This is 
incompatible with the ${\sim}1\,\mu$s per-cycle budget of superconducting 
qubits; a streaming decoder architecture (left to future work) would be 
required for real-time operation. The throughput numbers in 
Table~\ref{tab:speed} are appropriate for offline evaluation, simulation, 
and syndrome-buffered decoding scenarios.

AlphaQubit~2's real-time variant (AQ2-RT)~\cite{alphaqubit2} achieves 
sub-microsecond per-cycle throughput, placing it in the superconducting 
regime. AQ2-RT runs on Google's Trillium TPU 
accelerators~\cite{vahdat2024trillium}and employs hardware-specific 
optimizations including a streaming decoder architecture. The full AQ2 model, which runs in the ${\sim}10\,\mu$s 
regime on the same TPU hardware, is closer in throughput class to SMD's 
batched inference. Our goal in this work is not to match AQ2-RT's 
specialized-hardware performance, but to demonstrate that 
high-accuracy neural decoding approaching near-MLD performance can 
be achieved on a single commodity GPU with a small model 
($\leq 16$M parameters). The $\bigO(k)$ scaling of SMD is orthogonal to 
the streaming architecture of AQ2-RT---both optimizations could in 
principle be combined in future work to push sparse neural decoding into 
the sub-microsecond regime.

\begin{table}[t]
\centering
\caption{Effective input size at $p = 10^{-3}$ under SI1000 noise,
$R = 120$, $N = 10^5$ shots per cell.}
\label{tab:sparsity}
\small
\begin{tabular}{@{}lccc@{}}
\toprule
\textbf{Metric} & $d=3$ & $d=5$ & $d=7$ \\
\midrule
Stabilizers $(d^2-1)$                          & 8    & 24   & 48   \\
Dense input size $(d^2-1)\times R$             & 960  & 2880 & 5760 \\
Mean defects $k$ (measured)                    & 27.2 & 89.1 & 185  \\
99th-percentile $k$                            & 46   & 124  & 235  \\
Sparsity ratio $k / ((d^2 -1 ) R)$                   & 2.83\% & 3.09\% & 3.22\% \\
\bottomrule
\end{tabular}
\end{table}

\section{Discussion}
\label{sec:discussion}

\subsection{Sparse vs.\ Dense Mamba}
\label{sec:sparse_vs_dense}
 
The central comparison of this work is between SMD and the dense Mamba
decoder of Lee et al.~\cite{lee2025scalable}. Both use the same
underlying Mamba state-space model, but differ fundamentally in input
representation and spatial processing. The dense Mamba processes the
full $(d^2 - 1) \times R$ syndrome array recurrently, with each round's
stabilizer values passed through a stabilizer embedder and then through
Syndrome Mixer layers containing Mamba blocks, gated dense layers, and
\emph{dilated 2D convolutions} (dilation rates $[1, 1, 2]$ for $d = 5$)
that capture multi-scale spatial correlations on the surface code grid,
giving complexity $\bigO(d^2 R)$. SMD, in contrast, processes only $k$
active defects as a 1D sequence sorted by (time, stabilizer index). No
convolutions are used---spatial information is encoded entirely in the
13-dimensional feature vector (coordinates, neighbor flags, boundary
distances), giving complexity $\bigO(k)$.
 
On the Sycamore benchmark, our ensemble decoder achieves LERs of 
$2.940 \times 10^{-2}$ (distance~3) and $3.001 \times 10^{-2}$ 
(distance~5), matching or surpassing the dense Mamba's reported 
$2.98 \times 10^{-2}$ and $3.03 \times 10^{-2}$ 
(Figure~\ref{fig:sycamore}). This demonstrates that the sparse 
representation retains sufficient information for accurate decoding, 
even without the explicit 2D spatial processing provided by dilated 
convolutions. We attribute this result to two factors. First, the 
13-dimensional feature vector encodes rich spatial 
context---including normalized coordinates, stabilizer type, and binary 
neighbor flags---that implicitly captures much of the local geometry 
exploited by convolutions. Second, the Mamba selective scan mechanism 
learns to correlate defects across the sorted sequence, effectively 
discovering spatial relationships from the temporal ordering of defects 
sorted by (time, position).

On the uniform circuit-level noise benchmark, SMD outperforms MWPM 
by 12--53\% across $d = 3$--$7$ and the tested error-rate range, 
while maintaining nearly constant decoding latency (24--57\,$\mu$s) 
across $d = 3$--$9$. Under depolarizing noise 
with perfect stabilizers, the dual-head variant achieves 40--91\% lower 
error rates than MWPM, Tesseract, and Belief Matching at $d = 5$--$11$ 
by exploiting $Y$-error correlations through joint $(\lambda_Z, 
\lambda_X)$ decoding. The accuracy gap to matching decoders widens with 
both code distance and decreasing error rate, demonstrating that the 
sparse representation preserves the structural information needed for 
high-accuracy decoding.

\subsection{Feature Design Rationale}
\label{sec:feature_design}

The 13-dimensional defect representation encodes information analogous to what matching-based decoders rely on, while remaining compatible with variable-length sequence processing. The spatial coordinates 
$(x_i, y_i)$ and stabilizer type $\tau_i$ identify \emph{where} and 
\emph{what kind} of defect occurred---the same information that anchors each node in MWPM's matching graph. The boundary distances $b_Z, b_X$ encode 
proximity to the logical operators, which directly determines whether a 
chain of errors produces a logical failure; matching decoders encode 
this implicitly through boundary nodes. The four spatial neighbor flags 
$n_1, \ldots, n_4$ and two temporal neighbor flags $n_5, n_6$ provide a 
local error \emph{fingerprint}: an isolated defect (all flags zero) 
likely results from a measurement error, while a defect with fired 
neighbors signals a data qubit error---a distinction that matching 
decoders make through edge weights in the detector error model. Finally, 
the reconstructed measurement $m_{i,t}$ via cumulative XOR provides the 
model with both a \emph{change} signal (the detection event itself) and 
a \emph{state} signal (the current stabilizer value), following the 
dual-input approach of AlphaQubit~\cite{bausch2024learning}. Together, these features make the geometric structure of the surface code accessible to a purely sequential model, without requiring the learned 2D convolutions that dense decoders rely on to recover spatial context.

\subsection{Scalability and Limitations}
\label{sec:limitations}

The sparse approach offers several advantages. Computational cost
scales with the number of errors rather than the code size; geometry
is encoded directly in the feature vector, eliminating the need for
2D convolutions or spatial embedding; and training and inference run
on commodity GPUs without specialized hardware. The result is an
${\sim}7.5$--$16$M parameter decoder that achieves competitive accuracy
with $\bigO(k)$ scaling on a single RTX 4090 or H200 NVL.

Several extensions remain natural directions for future work. Our
compute budget restricted experimental evaluation to $d \leq 7$ under
SI1000 noise; with additional resources, the same training protocol
applies directly to $d = 9, 11$ and beyond. The present architecture
also processes the full $R$-round volume in one pass and is therefore
optimized for batched throughput (Table~\ref{tab:speed}) rather than
the ${\sim}1\,\mu$s per-cycle budget required for real-time decoding
on superconducting qubits. Reaching that regime would combine SMD's
$\bigO(k)$ scaling with the streaming, hardware-specific optimizations
developed for AQ2-RT~\cite{alphaqubit2}---a complementary direction
enabled by the small model size and sparse input representation.

This work focuses specifically on the rotated surface code, evaluated
on synthetic noise (depolarizing, uniform circuit-level, SI1000) and
on Sycamore experimental data. Natural extensions include applying
the sparse representation to recent surface-code experiments such as
Google's Willow distance-7 chip~\cite{google2024willow}, where the
same three-stage SI1000$\to$DEM$\to$experimental curriculum should
transfer directly; adapting the 13-dimensional features for color
codes and quantum LDPC codes, where structurally sparser syndromes
should make the $\bigO(k)$ scaling advantage even more pronounced;
implementing a streaming variant for real-time decoding; and
incorporating soft information (I/Q readout data) for further accuracy
gains.
\section{Conclusion}
\label{sec:conclusion}

We have introduced the Sparse Mamba Decoder, a neural decoder for 
quantum error correction that processes only active detection events 
rather than the full dense syndrome array. By representing each defect 
with 13 engineered features---including spatial coordinates, neighborhood 
connectivity, boundary distances, and reconstructed measurements via 
cumulative XOR---and processing the resulting variable-length sequences 
with a Mamba state-space model, our approach achieves $\bigO(k)$ 
computational complexity where $k$ is the number of defects.

Across four benchmarks spanning synthetic noise models and experimental 
data from Google's Sycamore quantum processor, SMD demonstrates 
competitive or superior accuracy compared to existing decoders. Under 
depolarizing noise with perfect stabilizers, the dual-head variant 
outperforms MWPM, Tesseract (near-MLD), and Belief Matching by 40--91\% 
at $d = 5$--$11$ through joint $(\lambda_Z, \lambda_X)$ decoding that 
exploits $Y$-error correlations---a structural advantage over 
matching-based decoders that decode each basis independently. Under uniform 
circuit-level noise, SMD outperforms MWPM at all distances 
$d = 3$--$9$. On the SI1000 benchmark, it reduces the MWPM 
logical error rate by up to 49\%, and on the Sycamore experimental 
benchmark, the ensemble matches or slightly surpasses the dense Mamba 
decoder of Lee et al.\ despite using a sparse input representation.

A key practical advantage is the decoder's favorable scaling: under 
uniform circuit-level noise at $p = 0.002$, latency increases only 2.4$\times$ from $d = 3$ to $d = 9$ (24\,$\mu$s to 57\,$\mu$s on RTX 4090), compared to 773$\times$ for Belief Matching over the same range. At $d = 9$, SMD decodes 230$\times$ faster than Belief Matching, while achieving higher accuracy than MWPM. Under SI1000 
noise, the speedup over near-MLD decoders grows to 
95--467$\times$ (Tesseract) and 232--463$\times$ (Belief Matching). 
With ${\sim}7.5$--$16$M parameters on commodity NVIDIA GPUs (RTX 4090 
for smaller models, H200 for the largest), SMD provides a practical 
alternative to approaches requiring specialized accelerator hardware.

Our results suggest that the inherent sparsity of quantum error 
syndromes at physically relevant error rates is a powerful structural 
prior that should be exploited by decoder architectures. As quantum 
processors scale to larger code distances, the gap between dense and 
sparse processing costs will only grow, making defect-centric 
approaches increasingly attractive. Future work includes adapting the 
sparse representation for color codes and quantum LDPC codes, 
implementing a streaming variant for real-time decoding, and combining 
the $\bigO(k)$ scaling with hardware-specific optimizations to push 
sparse neural decoding into the sub-microsecond regime.

\bibliographystyle{unsrtnat}
\bibliography{references}

\end{document}